\documentclass[11pt,a4paper]{article}

\usepackage{amsmath,amssymb,cite}
\usepackage{epsfig,graphicx} 

\begin{document}
\begin{flushright}
{\large DESY 06-187\\
Oktober 2006}
\end{flushright} 
%\vskip 2em
\vspace{2.0truecm}
\begin{center}
\boldmath
\large\bf \boldmath{$B \to (\rho,\omega)\gamma$} Decays and CKM Phenomenology
 \unboldmath
\end{center}

\vspace{0.9truecm}
\begin{center}
\bf Ahmed Ali\\[0.1cm]
{\sl Theory Group, Deutsches Elektronen-Synchrotron
 DESY,\\ Notkestrasse 85, 22603 Hamburg, Germany}\\[0.3cm]
\bf Alexander Parkhomenko\\[0.1cm]
{\sl Yaroslavl State (Demidov) University,\\ 
  Sovietskaya 14, 150000 Yaroslavl, Russia}
\end{center}

\vspace{8truecm}

\begin{center}
{\bf \sl Presented at the 14th International Seminar on High Energy
 Physics: Quarks-2006,\\St.~Petersburg, Russian Federation, 19 -- 25 May, 2006.\\
To appear in the Proceedings (Publishers: Institute for Nuclear Research, Moscow)}
\end{center}

\vfill
\setcounter{page}{0}
\thispagestyle{empty}
\newpage
\title{\bf
\boldmath{$B \to (\rho,\omega)\gamma$} Decays %\\  
and CKM Phenomenology  
}

\author{
Ahmed Ali$^a$\thanks{{\bf e-mail}: ahmed.ali@desy.de} 
and 
Alexander Parkhomenko$^b$\thanks{{\bf e-mail}: parkh@uniyar.ac.ru}\\  
$^a$ {\small\em Deutsches Elektronen-Syn\-chro\-tron DESY,} %\\ 
     {\small\em Notkestrasse 85, 22603 Hamburg, Germany} \\ 
$^b$ {\small\em Yaroslavl State (Demidov) University,} %\\ 
     {\small\em Sovietskaya 14, 150000 Yaroslavl, Russia}
}

\date{\today}

\maketitle

%\pagenumbering{arabic}   
\begin{abstract} 
We review and update the branching ratios for the 
$B \to (\rho,\omega) \gamma$ decays, calculated in 
the QCD factorization approach in the next-to-leading 
order (NLO) in the strong coupling~$\alpha_s$ and to 
leading power in~$\Lambda_{\rm QCD}/m_b$.
The corrections take into account the vertex, hard-spectator 
and annihilation contributions and are found to be large. 
Theoretical expectations for the branching ratios, 
CP-asymmetry, isospin- and $SU (3)_{\rm F}$-violating 
ratios in the $B \to \rho\gamma$ and $B \to \omega\gamma$ 
decays are presented and compared with the available data.
\end{abstract}

\section{Introduction} 

There is considerable theoretical interest in radiative 
$B \to V \gamma$ decays, where~$V$ is a vector meson 
($V = K^*, \, \rho, \, \omega, \, \phi$), 
as these processes are currently under intensive investigations 
in experiments at the two B-factories, BABAR and BELLE.    
The present measurements of the branching ratios 
for $B \to K^* \gamma$ decays from the CLEO~\cite{Coan:1999kh}, 
BABAR~\cite{Aubert:2004te}, and BELLE~\cite{Nakao:2004th} 
collaborations as well as their world averages~\cite{HFAG} 
are presented in Table~\ref{tab:exp-data}. In getting the 
isospin-averaged $B \to K^* \gamma$ branching fraction the 
following life-time weighted definition is adopted:  
\begin{equation}
\bar {\cal B} (B \to K^* \gamma) \equiv
\frac{1}{2} \left [ 
{\cal B} (B^+ \to K^{*+} \gamma) + 
\frac{\tau_{B^+}}{\tau_{B^{0}}} \, 
{\cal B} (B^0 \to K^{*0} \gamma) 
\right ] , 
\label{eq:BR-Ks-average-def}   
\end{equation}  
and the current world average~\cite{HFAG} 
for the $B$-meson lifetime ratio: 
\begin{equation} 
\tau_{B^+}/\tau_{B^0} = 1.076 \pm 0.008 , 
\label{eq:tauB+/tauB0-exp} 
\end{equation}
has been used in arriving at the numerical results.

\begin{table}[t] 
\caption{
Status of $B$-meson radiative branching fractions 
(in units of $10^{-6}$) after the ICHEP-2006 Conference 
(Moscow).  
} 
\label{tab:exp-data} 
\begin{center}
\begin{tabular}{l|ccc|c} 
\hline 
Mode & BABAR & BELLE & CLEO & HFAG 
\\ \hline
$B^+ \to K^{*+} \gamma$ & 
$38.7 \pm 2.8 \pm 2.6$ & 
$42.5 \pm 3.1 \pm 2.4$ & 
$37.6^{+8.9}_{-8.3} \pm 2.8$ & 
$40.3 \pm 2.6$ \\ 
$B^0 \to K^{*0} \gamma$ & 
$39.2 \pm 2.0 \pm 2.4$ & 
$40.1 \pm 2.1 \pm 1.7$ & 
$45.5^{+7.2}_{-6.8} \pm 3.4$ & 
$40.1 \pm 2.0$ \\ \hline 
$B^+ \to \rho^+ \gamma$ & 
$1.06^{+0.35}_{-0.31} \pm 0.09$ &
$0.55^{+ 0.42 + 0.09}_{- 0.36 - 0.08}$ & 
$< 13$ & 
$0.87^{+0.27}_{-0.25}$ \\ 
$B^0 \to \rho^0 \gamma$ & 
$0.77^{+0.21}_{-0.19} \pm 0.07$ &
$1.25^{+ 0.37 + 0.07}_{- 0.33 - 0.06}$ & 
$< 17$ & 
$0.91^{+0.19}_{-0.18}$ \\  
$B^0 \to \omega \gamma$ & 
$0.39^{+0.24}_{-0.20} \pm 0.03$ &
$0.56^{+ 0.34 + 0.05}_{- 0.27 - 0.10}$ & 
$< 9.2$ & 
$0.45^{+0.20}_{-0.17}$ \\ \hline 
$b \to s \gamma$ & 
$327 \pm 18^{+55}_{-41}$ & 
$355 \pm 32^{+30 +11}_{-31 -7}$ & 
$321 \pm 43^{+32}_{-29}$ & 
$355 \pm 24^{+9}_{-10} \pm 3$ \\  
$B \to K^* \gamma$ & 
$40.4 \pm 2.5$ & 
$42.8 \pm 2.4$ & 
$43.3 \pm 6.2$ & 
$41.8 \pm 1.7$ \\ 
$B \to (\rho,\omega) \, \gamma$ & 
$1.01 \pm 0.21 \pm 0.08$ &
$1.32^{+0.34 +0.10}_{-0.31 -0.09}$ & 
$< 14$ & 
$1.11^{+0.19}_{-0.18}$ \\ \hline 
\end{tabular} 
\end{center} 
\end{table}

The decays $B \to \rho \gamma$ and $B \to \omega \gamma$
have been experimentally searched since a long time,
as they are a measure of the underlying quark transition
$b \to d \gamma$. Hence, in the standard model (SM), 
they provide information on the Cabibbo-Kobayashi-Maskawa 
(CKM) matrix element~$|V_{td}|$. In particular, 
the ratio of the branching ratios
${\cal B}[B \to (\rho,\omega)\gamma]/{\cal B}(B \to K^*\gamma)$
provides an independent measurement of the CKM matrix 
element ratio $|V_{td}/V_{ts}|$, to be compared with the
corresponding ratio obtained through the ratio of the 
mixing-induced mass differences $\Delta M_{B_d}/\Delta M_{B_s}$, 
yielding~\cite{:2006ze} $|V_{td}/V_{ts}| = 
0.2060 ^{+0.0081}_{-0.0060} ({\rm theory}) \pm 0.0007 ({\rm exp})$. 

The first observation of the $B \to (\rho,\omega)\gamma$ decays 
was announced by the BELLE collaboration last summer~\cite{Abe:2005rj}, 
and the results are presented in Table~\ref{tab:exp-data}. Of these, 
the signal from the $B^0 \to \rho^0 \gamma$ decay was established
with a significance of~$5.2\sigma$  while no evidence from the other 
two decay modes $B^+ \to \rho^+ \gamma$ and $B^0 \to \omega \gamma$ 
was found (their significances are~$1.6\sigma$ and~$2.3\sigma$, 
respectively). The branching fraction of the 
charged mode $B^+ \to \rho^+ \gamma$  is currently a factor two smaller 
than that of the neutral decay mode $B^0 \to \rho^0 \gamma$~--
in obvious contradiction with the SM 
predictions~\cite{Ali:2001ez,Ali:2004hn}.
However, one should not try to read too much from 
the existing data which are statistically limited.

At the ICHEP-2006 Conference in Moscow this summer~\cite{unknown:2006ag},
the BABAR collaboration have also presented the measurements 
of the $B \to \rho \gamma$ and $B \to \omega \gamma$ branching 
fractions, which are shown in Table~\ref{tab:exp-data}.  
Based on approximately the same statistics as the BELLE collaboration, 
in the BABAR data both the charged and neutral $B \to \rho \gamma$ decays
were observed  
with the significances~$4.1\sigma$ and~$5.2\sigma$, respectively. 
There is no evidence for the $B^0 \to \omega \gamma$ decay mode  
yet (the signal has a significance of~$2.3\sigma$). Thus, both the
 collaborations have 
 observed the $B^0 \to \rho^0 \gamma$ mode in  good agreement 
with each other within the experimental errors,  while the other two decay
modes require more statistics to be established.        
With limited statistics, one may resort to the following 
weighted branching fraction for the CKM phenomenology: 
\begin{equation} 
{\cal B} [B \to (\rho,\omega) \, \gamma] \equiv
\frac{1}{2} \left \{ {\cal B} (B^+ \to \rho^+ \gamma) 
+ \frac{\tau_{B^+}}{\tau_{B^0}} \left [ 
{\cal B} (B^0 \to \rho^0 \gamma) + 
{\cal B} (B^0 \to \omega \gamma) \right]   
\right \} .  
\label{eq:BR-average-def}   
\end{equation}
Both the BABAR and BELLE collaborations have measured
this fraction with ~$5.1\sigma$ significance 
(see Table~\ref{tab:exp-data}) and within errors their 
measurements agree.

The other potentially interesting radiative mode is the decay
$B^0 \to \phi\gamma$. Dominated by the anni\-hi\-la\-tion-type 
diagrams, its branching fraction has been estimated at the 
level of~$10^{-11}$~\cite{Li:2006xe,Lu:2006nz}, 
too small to be measured at present $B$-meson factories, 
but, possibly this mode can be targeted by the LHC-b 
experiment or at a future high-luminosity Super-B factory. 
The current upper limit on this decay (at 90\% C.L.) 
is reported by the BABAR collaboration~\cite{Aubert:2005qc}: 
\begin{equation} 
{\cal B}_{\rm exp} (B^0 \to \phi\gamma) < 0.85 \times 10^{-6} . 
\label{eq:B-phi-gamma-exp} 
\end{equation} 
No information on this decay from the BELLE collaboration is 
as yet available.

What concerns the CKM phenomenology, the ratios of the 
branching fractions are more reliably calculable, as the 
various uncertainties related to the theoretical input 
are considerably reduced in these ratios thereby enhancing 
the precision on the ratio $\vert V_{td}/V_{ts}\vert$. 
One such ratio is defined below together with its current 
experimental measurements: 
\begin{equation} 
R_{\rm exp} [(\rho,\omega) \gamma/K^* \gamma] \equiv 
\frac{{\cal B}_{\rm exp} [B \to (\rho,\omega) \gamma]}
     {{\cal B}_{\rm exp} (B \to K^* \gamma)} = 
\left \{ 
\begin{array}{ll} 
0.024 \pm 0.005 ,           & [{\rm BABAR}] \\ 
0.032 \pm 0.008 \pm 0.002 . & [{\rm BELLE}] 
\end{array}
\right. 
\label{eq:ratio-exp}
\end{equation}
The results presented are consistent with each other 
within errors. 

Ratios of neutral $B$-meson branching fractions are more 
favorable for the CKM analysis as they are less sensitive 
to the annihilation contribution, which is theoretically
less tractable but expected to be small for the neutral modes.
The BABAR collaboration have presented the measurement 
of such a ratio~\cite{unknown:2006ag}:   
\begin{equation} 
R_{\rm exp} (\rho^0 \gamma/K^{*0} \gamma) \equiv 
\frac{2 \, {\cal B}_{\rm exp} (B^0 \to \rho^0 \gamma)}
     {{\cal B}_{\rm exp} (B^0 \to K^{*0} \gamma)} = 
0.038^{+0.011}_{-0.010} . 
\label{eq:ratio-neutral-exp}
\end{equation}
In comparison with Eq.~(\ref{eq:ratio-exp}), the central value 
in Eq.~(\ref{eq:ratio-neutral-exp}) is substantially larger 
but due to the large errors the two measurements are compatible 
with each other. As emphasized by several authors in the past,
measurements of these ratios provide a robust determination 
of the ratio~$|V_{td}/V_{ts}|$ of the CKM matrix elements. 
However, to make an impact on the CKM phenomenology, in particular 
in the post-$\Delta M_s$ observation era, the measurements in 
radiative $B$-meson decays have to become an order of magnitude 
more precise than is currently the case. In view of this, we will 
constrain the CKM parameters from the SM fits of the unitarity 
triangle~\cite{Charles:2004jd,Bona:2006ah}, including the measurement
of $\Delta M_s$~\cite{:2006ze},
 and predict the various branching ratios, their ratios, 
and asymmetries to be confronted with data in radiative $B$-decays. 
This will serve as a stringent test of the SM in this sector.

Several competing theoretical frameworks have been used 
to study exclusive $B$-meson decays. The QCD-Factorization 
approach~\cite{Beneke:1999br} provides a satisfactory 
theoretical basis for calculations of two-body radiative 
$B$-meson decays~\cite{Beneke:2000wa} and has been applied 
to the $B \to K^*\gamma$~\cite{%
Ali:2001ez,Beneke:2001at,Bosch:2001gv,Kagan:2001zk,%
Ali:2004hn,Bosch:2004nd,Beneke:2004dp}, 
$B \to \rho\gamma$~\cite{Ali:2001ez,Ali:2004hn,%
Bosch:2001gv,Ali:2002kw,Bosch:2004nd,Beneke:2004dp} 
and $B \to \phi\gamma$~\cite{Li:2003kz} modes. 
There are several other theoretical approaches which have 
also been used to study two-body radiative $B$-meson decays. 
These include the Soft-Collinear Effective Theory 
(SCET)~\cite{Chay:2003kb,Becher:2005fg,Lu:2006nz} 
and the perturbative QCD (pQCD) 
approach~\cite{Keum:2004is,Lu:2005yz,Li:2006xe}.
In addition, information on various input hadronic quantities
is required which is usually taken 
from the Light-Cone Sum Rules 
(LCSRs)~\cite{Ball:2006nr,Ball:2006cv}.  
All these approaches are in fair agreement with 
the measured branching ratios of the $B \to K^* \gamma$ decays, 
and predict the branching ratios of the $B \to \rho\gamma$ 
and $B \to \omega\gamma$ decays typically of $O(10^{-6})$. 

In this paper, we discuss and review the predictions for the 
branching ratio of the $B \to \rho \gamma$ and $B \to \omega\gamma$ 
decays obtained in the QCD-Factorization framework. 
We shall concentrate mainly on the ratio of the branching 
fractions defined below: 
\begin{equation}
R_{\rm th}(\rho \gamma/K^* \gamma) \equiv
\frac{{\cal B}_{\rm th} (B \to \rho \gamma)}
     {S_\rho \, {\cal B}_{\rm th} (B \to K^* \gamma)} , 
\qquad 
R_{\rm th}(\omega \gamma/K^* \gamma) \equiv 
\frac{2 \, {\cal B}_{\rm th} (B \to \omega \gamma)}
     {{\cal B}_{\rm th} (B \to K^* \gamma)} ,  
\label{eq:ratio_th} 
\end{equation} 
where $S_\rho = 1$ for the $B^\pm$-meson decay modes 
and $S_\rho = 1/2$ for the $B^0$-meson decays.  
Measurements of the $B \to K^* \gamma$ branching ratios 
in combination with the theoretical estimates of the ratios 
in~(\ref{eq:ratio_th}) allow us to make predictions for the
$B \to \rho \gamma$ and $B \to \omega\gamma$ branching fractions 
with reduced uncertainties.

In addition to the branching ratios, there are several asymmetries
involving isospin-, $SU(3)_{\rm F}$- and CP-violation in the 
$B \to (\rho, \omega) \gamma$ decays. For example, first 
measurement of the isospin-violating ratio, defined below, has been 
presented by the BABAR collaboration this summer~\cite{unknown:2006ag}: 
\begin{equation} 
\Delta \equiv \frac{1}{2} \, 
\frac{\Gamma (B^+ \to \rho^+\gamma)}
     {\Gamma (B^0 \to \rho^0\gamma)} - 1  
= \frac{\tau_{B^0}}{2 \, \tau_{B^+}} \, 
\frac{{\cal B} (B^+ \to \rho^+\gamma)}
     {{\cal B} (B^0 \to \rho^0\gamma)} - 1 = 
-0.36 \pm 0.27 , 
\label{eq:Delta-exp}
\end{equation}
which is  consistent with zero at $1.3\sigma$. 
As the isospin-violating ratio~$\Delta$ depends on the 
unitarity-triangle angle~$\alpha$ due to the interference 
between the penguin- and annihilation-type contributions 
 (see, Eq.~(\ref{eq:Delta-LO}) below), 
its experimental measurement, in principle, will yield 
an independent determination of this angle. In the SM, 
constraining the angle~$\alpha$ from the unitarity fits,
$\alpha = (97.3^{+4.5}_{-5.0})^\circ$~\cite{Charles:2004jd}, 
we estimate $\Delta =(2.9 \pm 2.1)\%$ in the QCD factorization 
approach. Estimates of $\Delta$ in the pQCD approach~\cite{Li:2006xe}
are similar though they allow somewhat larger isospin-violation
Thus,  isospin-violation in the $B \to \rho\gamma$ 
decays is parametrically small in the SM, 
being a consequence of the experimentally measured value 
$|\cos\alpha|_{\rm SM} <0.2$ and the ratio of the 
annihilation-to-penguin amplitudes, typically estimated as $|A/P| \leq 0.3$. 
The $SU(3)_{\rm F}$-violating ratio $\Delta^{(\rho/\omega)}$, 
defined in Eq.~(\ref{eq:SU3-ratio}), is estimated to be 
likewise small in the SM. With a realistic estimate of the
$SU(3)_{\rm F}$-breaking in the form factors,  
 $\zeta_{\omega/\rho} \equiv 
\xi_\perp^{(\omega)}(0)/\xi_\perp^{(\rho)}(0)= 0.9 \pm 0.1$, 
we estimate  $\Delta^{(\rho/\omega)} = (11 \pm 11)\%$, which is
 consistent with the current data within large experimental errors. 
These predictions can be tested in high statistics measurements
in the $B \to (\rho,\omega) \gamma$ decays. Finally, we also 
present the CP-asymmetries (both direct and mixing-induced) 
in the $B \to \rho\gamma$ and $B \to \omega \gamma$ decays,
updating our results presented 
in Refs.~\cite{Ali:2001ez,Ali:2004hn}.  These asymmetries 
test the underlying dynamical model (the QCD factorization), 
as shown by comparison with the corresponding existing 
calculations in the pQCD approach~\cite{Lu:2005yz}.

\section{\boldmath{$B \to V \gamma$} Branching Fractions in NLO}

The effective Hamiltonian for the $B \to \rho \gamma$  
(equivalently $b \to d \gamma$) decays at the scale 
$\mu = O (m_b)$, where~$m_b$ is the $b$-quark mass, 
is as follows: 
\begin{eqnarray}
{\cal H}_{\rm eff} & = & \frac{G_F}{\sqrt 2} 
%\, \sum_{p = d, s}
\left \{
V_{ub} V_{ud}^* \,
\left [
C_1 (\mu) \, {\cal O}_1^{(u)} (\mu) +
C_2 (\mu) \, {\cal O}_2^{(u)} (\mu)
\right ]
\right.
\nonumber \\
& + & 
V_{cb} V_{cd}^* \,
\left [
C_1 (\mu) \, {\cal O}_1^{(c)} (\mu) +
C_2 (\mu) \, {\cal O}_2^{(c)} (\mu)
\right ]
\label{eq:Hamiltonian} \\
& - &   
\left.
V_{tb} V_{td}^* \,
\left [
C_7^{\rm eff} (\mu) \, {\cal O}_{7\gamma} (\mu) +
C_8^{\rm eff} (\mu) \, {\cal O}_{8g} (\mu)
\right ]
+ \ldots
\right \} ,
\nonumber
\end{eqnarray}
where the set of operators is ($q = u, c$): 
\begin{eqnarray}
{\cal O}_1^{(q)} & = & 
(\bar d_\alpha \gamma_\mu (1 - \gamma_5) q_\beta) \,
(\bar q_\beta \gamma^\mu (1 - \gamma_5) b_\alpha) ,
\label{eq:operator-O1} \\  
{\cal O}_2^{(q)} & = & 
(\bar d_\alpha \gamma_\mu (1 - \gamma_5) q_\alpha) \,
(\bar q_\beta \gamma^\mu (1 - \gamma_5) b_\beta) ,
\label{eq:operator-O2} \\  
{\cal O}_{7\gamma} (\mu) & = & \frac{e \, m_b (\mu)}{8 \pi^2} \,
(\bar d_\alpha \sigma^{\mu \nu} (1 + \gamma_5) b_\alpha) \,
F_{\mu \nu} ,   
\label{eq:operator-O7} \\  
{\cal O}_{8g} (\mu) & = & \frac{g_s (\mu) \, m_b (\mu)}{8 \pi^2} \,
(\bar d_\alpha \sigma^{\mu \nu} (1 + \gamma_5)
T^A_{\alpha \beta} b_\beta) \, G^A_{\mu \nu} .
\label{eq:operator-O8}    
\end{eqnarray}
The strong and electroweak four-quark penguin operators 
are present in the effective Hamiltonian (denoted 
by ellipses) but are not taken into account due to 
their small Wilson coefficients. 

The effective Hamiltonian sandwiched between the~$B$-meson 
and the vector meson~$V$ states can be expressed in terms 
of the matrix elements of bilinear quark currents defining 
a heavy-to-light transition.
% These matrix elements are dominated 
%by strong interactions at small momentum transfer and cannot 
%be calculated perturbatively.
 The general decomposition of 
the matrix elements on all possible Lorentz structures admits 
seven scalar functions (form factors): $V^{(V)} (q^2)$, 
$A_i^{(V)} (q^2)$ ($i = 0, 1, 2$), and~$T_i^{(V)} (q^2)$ 
($i = 1, 2, 3$) of the momentum squared~$q^2 = (p_B - p)^2$ 
transferred from the $B$-meson to the light vector meson. 
To be definite, we study the $B \to \rho \gamma$ decay in 
which the transition matrix elements are defined as follows: 
\begin{eqnarray}
&&
\left < \rho (p, \varepsilon^*) | 
\bar d \, \gamma^\mu \, b 
| \bar B (p_B) \right > = 
\frac{2 i \, V^{(\rho)} (q^2)} {m_B + m_\rho} \, 
\varepsilon^{\mu \nu \alpha \beta} \,  
\varepsilon^*_\nu p_\alpha p_{B \beta} , 
% {\rm eps} (\mu, \varepsilon^*, p, p_B) ,
\label{eq:vertor-current} \\
&& 
\left < \rho (p, \varepsilon^*) | 
\bar d \, \gamma^\mu \gamma_5 \, b 
| \bar B (p_B) \right > = 
% 2 m_\rho \, A_0^{(\rho)} (q^2) \, 
% \frac{(\varepsilon^* q)}{q^2} \, q^\mu
% \label{eq:axial-current} \\ 
% && 
% \hspace{25mm}
% + 
A_1^{(\rho)} (q^2) \, (m_B + m_\rho) 
\left [ \varepsilon^{* \mu} - 
\frac{(\varepsilon^* q)}{q^2} \, q^\mu \right ] 
\label{eq:axial-current} \\ 
% \nonumber \\ 
&& 
\hspace{25mm}  
- A_2^{(\rho)} (q^2) \, \frac{(\varepsilon^* q)}{m_B + m_\rho} \, 
\left [ (p_B + p)^\mu - \frac{(m_B^2 - m_\rho^2)}{q^2} q^\mu  
\right ] %,
+ 2 m_\rho \, A_0^{(\rho)} (q^2) \, 
\frac{(\varepsilon^* q)}{q^2} \, q^\mu ,  
\nonumber \\
&&
\left < \rho (p, \varepsilon^*) | 
\bar d \, \sigma^{\mu \nu} q_\nu \, b
| \bar B (p_B) \right > = 
2 \, T_1^{(\rho)} (q^2) \, 
\varepsilon^{\mu \nu \alpha \beta}  
\varepsilon^*_\nu p_\alpha p_{B \beta} , 
% {\rm eps} (\mu, \varepsilon^*, p, p_B) ,
\label{eq:tensor-current} \\
&& 
\left < \rho (p, \varepsilon^*) | 
\bar d \, \sigma^{\mu \nu} \gamma_5 q_\nu \, b 
| \bar B (p_B) \right >
% \label{eq:axial-tensor-current} \\ 
% && 
% \hspace{25mm} 
= - i \, T_2^{(\rho)} (q^2) \, [(m_B^2 - m_\rho^2) \, 
\varepsilon^{* \mu} - (\varepsilon^* q) \, (p_B + p)^\mu]
\label{eq:axial-tensor-current} \\
% \nonumber \\
&& 
\hspace{30mm} 
- i \, T_3^{(\rho)} (q^2) \, (\varepsilon^* q) \,
\left [
q^\mu - \frac{q^2}{m_B^2 - m_\rho^2} \, (p_B + p)^\mu
\right ] .   
\nonumber 
\end{eqnarray}
%
% where ${\rm eps} (\mu, \varepsilon^*, p, p_B) = 
% \varepsilon^{\mu \nu \alpha \beta}  
% \varepsilon^*_\nu p_\alpha p_{B \beta}$. 
The heavy quark symmetry in the large energy limit of the vector 
meson allows to reduce the number of independent form factors 
to two only: $\xi^{(\rho)}_\perp (q^2)$ and $\xi^{(\rho)}_\| (q^2)$.
Both of them enter in the analysis of the $B \to \rho \ell^+\ell^-$
decay. However, for the radiative $B \to \rho \gamma$ decay amplitude, 
we need only one of them, $\xi^{(\rho)}_\perp (q^2=0)$, 
which is related to the form factors introduced above in the full 
QCD as follows (terms of order $m_\rho^2/m_B^2$ are neglected): 
\begin{equation} 
\frac{m_B}{m_B + m_\rho} \, V^{(\rho)} (0) = 
\frac{m_B + m_\rho}{m_B} \, A_1^{(\rho)} (0) = 
T_1^{(\rho)} (0) = T_2^{(\rho)} (0) = \xi^{(\rho)}_\perp (0) .  
\label{eq:FF-relation}
\end{equation}  
These relations among the form factors in the symmetry 
limit are broken by perturbative QCD radiative corrections
arising from the vertex renormalization and hard-spectator
interaction.
To incorporate both types of QCD corrections, a
factorization formula for the heavy-to-light transition form 
factors at large recoil and at leading order in the inverse 
heavy meson mass was established in Ref.~\cite{Beneke:2000wa}:
\begin{equation} 
F^{(\rho)}_k (q^2 = 0) = 
C_{\perp k} \, \xi^{(\rho)}_\perp (q^2 = 0) + 
% C_{\| k} \xi^{(\rho)}_\| +
\phi_B \otimes T_k (q^2 = 0) \otimes \phi_\rho ,
\label{eq:factor-general}
\end{equation} 
where $F^{(\rho)}_k (q^2 = 0)$ is any of the four form factors in
the $B \to \rho$ transitions related by Eq.~(\ref{eq:FF-relation}), 
$C_{\perp k} = C_{\perp k}^{(0)} \, [ 1 + O (\alpha_s)]$ is the  
renormalization coefficient, $T_k$ is a hard-scattering kernel 
calculated in $O (\alpha_s)$,  
$\phi_B$ and~$\phi_\rho$ are the light-cone distribution  
amplitudes (LCDAs) of the~$B$- and $\rho$-mesons convoluted with the 
kernel~$T_k$. 

In the leading order, the electromagnetic penguin 
operator~${\cal O}_{7\gamma}$ contributes in the 
$B \to \rho \gamma$ decay amplitude at the tree level. 
Taking into account the definitions of the $B \to \rho$ 
transition form factors in the tensor~(\ref{eq:tensor-current}) 
and the axial-tensor~(\ref{eq:axial-tensor-current}) currents 
and the symmetry relation $T^{(\rho)}_1 (0) = T^{(\rho)}_2 (0)$, 
the amplitude for the $B \to \rho \gamma$ decay takes the form: 
\begin{eqnarray}
M^{(0)} & = & - \frac{G_F}{\sqrt 2} \, V_{tb} V_{td}^* \, 
\frac{e \bar m_b (\mu)}{4 \pi^2} \, C_7^{(0) {\rm eff}} (\mu) \, 
T^{(\rho)}_1 (0) 
\label{eq:ME0} \\ 
& \times &
\left [ (P q) (e^* \varepsilon^*) - (e^* P) (\varepsilon^* q) 
+ i \, {\rm eps} (e^*, \varepsilon^*, P, q)   
\right ] ,
\nonumber 
\end{eqnarray}
where $q = p_B - p$ and $e^*$ are the photon four-momentum 
and polarization vector, respectively, $P = p_B + p$, and 
${\rm eps} (e^*, \varepsilon^*, P, q) = 
\varepsilon^{\mu \nu \alpha \beta} \,   
e^*_\mu \varepsilon^*_\nu P_\alpha q_\beta$.  
The corresponding branching ratio can be easily obtained and reads as follows: 
\begin{equation}
{\cal B}_{\rm th}^{\rm LO} (B \to \rho \gamma) = 
\tau_B \, S_\rho \, 
\frac{G_F^2 \alpha |V_{tb} V_{td}^*|^2 m_B^3}{32 \pi^4} \, 
\left [ 1 - \frac{m_\rho^2}{m_B^2} \right ]^3 
\bar m_b^2 (\mu) \, |C_7^{(0) {\rm eff}} (\mu)|^2 \, 
|T^{(\rho)}_1 (0, \mu)|^2 ,    
\label{eq:Br-rho}
\end{equation}
where $S_\rho = 1$ for the $B^\pm$-meson decay
and $S_\rho = 1/2$ for the $B^0$ decay. 
The scale ($\mu$)-dependence of the form factor, $T^{(\rho)}_1 (0, \mu)$, 
the $b$-quark mass, $\bar m_b (\mu)$, and the Wilson coefficient, 
$C_7^{(0) {\rm eff}} (\mu)$, in the above expression 
for the branching ratio are made explicit. 

The branching fraction for the $B \to K^* \gamma$ decays can be easily 
obtained from Eq.~(\ref{eq:Br-rho}) by replacing $V_{td} \to V_{ts}$, 
$m_\rho \to m_{K^*}$, and $T^{(\rho)}_1 (0, \mu) \to T^{(K^*)}_1 (0, \mu)$,  
which yields the following expression for the ratio of
the branching ratios defined in Eq.~(\ref{eq:ratio_th}): 
\begin{equation} 
R_{\rm th}^{(0)} (\rho \gamma/K^* \gamma) = 
S_\rho \left | \frac{V_{td}}{V_{ts}} \right |^2 
\left [ \frac{m_B^2 - m_\rho^2}{m_B^2 - m_{K^*}^2} \right ]^3 
\left [ \frac{T^{(\rho)}_1 (0, \mu)}{T^{(K^*)}_1 (0, \mu)} \right ]^2.    
\label{eq:ratio0_result}
\end{equation}
A similar ratio involving the $B^0 \to \omega \gamma$ and 
$B^0 \to K^{*0}\gamma$ decay widths can be written as follows: 
\begin{equation} 
R_{\rm th}^{(0)} (\omega \gamma/K^* \gamma) = 
\frac{1}{2} \, \left | \frac{V_{td}}{V_{ts}} \right |^2 
\left [ \frac{m_B^2 - m_\omega^2}{m_B^2 - m_{K^*}^2} \right ]^3 
\left [ \frac{T^{(\omega)}_1 (0, \mu)}{T^{(K^*)}_1 (0, \mu)} \right ]^2.    
\label{eq:ratio0_omega_result}
\end{equation}

Apart from the electromagnetic penguins, one also has 
contributions from the annihilation diagrams to the 
$B \to \rho \gamma$ and $B \to \omega \gamma$ decay 
widths which modify the ratios~(\ref{eq:ratio0_result}) 
and~(\ref{eq:ratio0_omega_result}): 
\begin{eqnarray}
R_{\rm th} (\rho \gamma/K^* \gamma) =
R_{\rm th}^{(0)} (\rho \gamma/K^* \gamma)
\left [ 1 + \Delta R (\rho/K^*) \right ] , 
\label{eq:ratio_ann_result} \\ 
R_{\rm th} (\omega \gamma/K^* \gamma) =
R_{\rm th}^{(0)} (\omega \gamma/K^* \gamma)
\left [ 1 + \Delta R (\omega/K^*) \right ] .
\label{eq:ratio_ann_omega_result}
\end{eqnarray}
In the annihilation amplitude, photon radiation from the quarks in the vector
 meson is 
compensated by the diagram in which  the photon is emitted from the 
vertex~\cite{Khodjamirian:1995uc,Ali:1995uy,Khodjamirian:2001ga}. 
Hence, only the annihilation diagram with the photon emitted from 
the spectator quark in the $B$-meson is numerically important.
The quantities $\Delta R(\rho/K^*)$ and $\Delta R(\omega/K^*)$  
 can be parameterized  (apart from the CKM factors) by  dimensionless 
factors~$\epsilon_A^{(\pm)}$, ~$\epsilon_A^{(0)}$ and
$\epsilon_A^{\omega}$:  
\begin{eqnarray}
&&
\Delta R (\rho^\pm/K^{*\pm}) = \lambda_u \, \varepsilon_A^{(\pm)} ,
\qquad 
\Delta R (\rho^0/K^{*0}) = \lambda_u \, \varepsilon_A^{(0)} ,
\qquad 
\Delta R (\omega/K^{*0}) = \lambda_u \, \varepsilon_A^{(\omega)} ,
\qquad 
\label{eq:ann_contribution} \\
&&
\hspace*{30mm} 
\lambda_u = \frac{V_{ub} V_{ud}^*}{V_{tb} V_{td}^*} =
- \left | \frac{V_{ub} V_{ud}^*}{V_{tb} V_{td}^*} \right |
{\rm e}^{i \alpha}
= F_1 + i F_2 , 
\label{eq:CKM-ratio}
\end{eqnarray}
where~$F_1=-|\lambda_u|\cos \alpha$, $F_2=-|\lambda_u|\sin \alpha$,
and $\alpha$ is one of the inner angles of the unitarity triangle.
In the neutral $B$-meson decays, the parameter~$\varepsilon_A$ 
is numerically small due to the color suppression and
the unfavorable electric charge of the $d$-quark, resulting in the
 estimate $\varepsilon_A^{(0)} = 
- \varepsilon_A^{(\omega)} = 0.03 \pm 0.01$~\cite{Ali:1995uy}, 
obtained with the help of the Light-Cone Sum Rules.  
For the charged $B$-meson decays, LCSRs yield a larger value 
$\varepsilon_A^{(\pm)} = 0.30 \pm 0.07$~\cite{Ali:1995uy}, which 
is used in the current analysis.  

Both the penguin and annihilation contributions receive 
QCD corrections. 
% They have been calculated for the penguin-amplitudes.
% Taking them into account results in the next-to-leading order 
% (NLO) expression for the decay widths.
The next-to-leading order (NLO) corrections to the 
$B \to \rho \gamma$ and $B \to \omega \gamma$ decay widths 
consist of the following contributions~\cite{Ali:2001ez}:  
\begin{enumerate} 
\item 
The NLO correction to the $\overline{\rm MS}$ $b$-quark mass 
$\bar m_b (\mu)$. We have related~$\bar m_b (\mu)$ with the pole 
mass, $m_{b, {\rm pole}}$, at the renormalization scale~$\mu$. 

\item 
The NLO correction to the Wilson coefficient $C_7^{\rm eff} (\mu)$. 

\item
The factorizable NLO corrections to the $T^{(\rho)}_1 (0, \mu)$  
and $T^{(\omega)}_1 (0, \mu)$ form factors which can be further 
divided into the vertex and hard-spectator corrections. 
These two types of corrections are estimated at different scales: 
the vertex and hard-spectator corrections should be calculated 
at the hard $\mu_b \sim m_b$ and intermediate $\mu_i \sim 
\sqrt{\Lambda_{\rm H} m_b}$ ($\Lambda_{\rm H} \simeq 0.5$~GeV) 
scales, respectively. 

\item
The nonfactorizable NLO corrections which are also of two types: 
the vertex and the hard-spectator corrections. The nonfactorizable 
vertex corrections can be taken from inclusive $B \to X_d \gamma$ 
decay~\cite{Ali:1998rr}. 
The nonfactorizable hard-spectator corrections were calculated 
by several groups~\cite{Ali:2001ez,Beneke:2001at,Bosch:2001gv}. 

\end{enumerate}
In addition, the NLO corrections to the annihilation diagrams
have also to be taken into account. We have mentioned them in 
the context of the $B \to \phi \gamma$ decay. For the 
$B^\pm \to \rho^\pm \gamma$ decay, they can be modeled on the 
$B^\pm \to \ell^\pm \nu_\ell \gamma$ decay, as based on the 
large-$N_c$ argument, the non-factorizing contribution is 
expected to be small~\cite{Grinstein:2000pc}. We shall adopt 
here the annihilation contribution estimates obtained using 
the QCD LCSRs~\cite{Ali:1995uy,Khodjamirian:2001ga}.
 
The NLO corrections discussed above modify the $B \to \rho\gamma$
and $B \to \omega\gamma$ branching ratios. The result for the 
charged-conjugate averaged $B^\pm \to \rho^\pm \gamma$ branching 
fraction can be written in the form:
\begin{eqnarray}
&& \hspace*{-10mm}
\bar {\cal B}_{\rm th} (B^\pm \to \rho^\pm \gamma) =
\tau_{B^+} \, \frac{G_F^2 \alpha |V_{tb} V_{td}^*|^2}{32 \pi^4} \,
m_{b,{\rm pole}}^2 \, m_B^3
\left [ 1 - \frac{m_\rho^2}{m_B^2} \right ]^3
\left [ \xi_\perp^{(\rho)} (0) \right ]^2 \, C^{(0){\rm eff}}_7 
\label{eq:DecayWidth} \\
&& \hspace*{-10mm}
\times
\left \{ C^{(0){\rm eff}}_7 + 2 A^{(1)t}_R +
\epsilon_A^{(\pm)} \, (F_1^2 + F_2^2) \, 
[ \epsilon_A^{(\pm)} \, C^{(0) {\rm eff}}_7 + 2 A^u_R ]
%\right.
%\nonumber \\
%& + &
%\left.
+ 2 F_1 \, [ A^u_R + \epsilon_A^{(\pm)} \, 
(C^{(0){\rm eff}}_7 + A^{(1)t}_R)]
%\mp 2 F_2 \, [ C^{(0){\rm eff}}_7 A^u_I - A^{(1)t}_I L^u_R ]
\right \} ,
\nonumber
\end{eqnarray}
where 
% $L^u_R = \epsilon_A \, C^{(0) {\rm eff}}_7$.
the subscript~$R$ denotes the real part of the corresponding quantity. 
The NLO amplitude~$A^{(1)t} (\mu)$ of the decay presented here 
can be decomposed in three contributing parts~\cite{Ali:2001ez}:
\begin{equation}
A^{(1)t} (\mu) = A^{(1)}_{C_7} (\mu) +
A^{(1)}_{\rm ver} (\mu) + A^{(1)\rho}_{\rm sp} (\mu_{\rm sp}), 
\label{eq:NLO-ampl-decomposition}
\end{equation}
where the correction due to the $b$-quark mass, $\bar m_b (\mu)$, 
is included in  $A^{(1)}_{\rm ver} (\mu)$. 
The amplitude~$A^{(1) K^*} (\mu)$ for the $B \to K^* \gamma$ decay
can be written in a similar form and differs from~$A^{(1)t} (\mu)$ 
by the hard-spectator part~$A^{(1) K^*}_{\rm sp} (\mu)$ 
only~\cite{Ali:2001ez}. 
Note that the $u$-quark contribution~$A^u (\mu)$ from the penguin 
diagrams, which also involves the contribution of hard-spectator 
corrections, can not be ignored in the $B \to \rho \gamma$ and 
$B \to \omega \gamma$ decays.

Using the formula~(\ref{eq:DecayWidth}) for the branching ratio,
the dynamical function $\Delta R (\rho/K^*)$, defined by
Eq.~(\ref{eq:ratio_ann_result}),  can be written
as follows~\cite{Ali:2001ez}:
\begin{eqnarray}
&& \Delta R (\rho/K^*) =
\left [ 2 \epsilon_A \, F_1 + \epsilon_A^2 (F_1^2 + F_2^2) \right ] 
\left ( 1 - \frac{2 A^{(1)K^*}}{C^{(0) {\rm eff}}_7} \right ) - 
\frac{2 A^{(1)K^*}}{C^{(0) {\rm eff}}_7}  
\label{eq:DR-NLO} \\
&&  \hspace*{10mm}
+ \frac{2}{C^{(0) {\rm eff}}_7} \, {\rm Re}
\left [
 A^{(1)\rho}_{\rm sp} -  A^{(1)K^*}_{\rm sp}
+ F_1 (A^u + \epsilon_A A^{(1)t}) + \epsilon_A (F_1^2 + F_2^2) A^u
\right ] ,
\nonumber
\end{eqnarray}
where the NLO corrections in the penguin amplitude and QCD LCSRs
for the annihilation amplitude are taken into account.
A similar expression with the exchange 
$\epsilon_A \to \epsilon_A^{(\omega)}$ 
holds for~$\Delta R (\omega/K^*)$ defined 
in~(\ref{eq:ratio_ann_omega_result}).

\section{
Phenomenology of \boldmath{$B \to \rho \gamma$} 
and \boldmath{$B \to \omega \gamma$} Decays
}

\paragraph{
Branching Ratios for \boldmath{$B \to \rho \gamma$} 
and \boldmath{$B \to \omega \gamma$} Decays:
} 
For the numerical predictions for the $B \to \rho \gamma$ 
and $B \to \omega \gamma$ branching ratios, we employ the ratios
defined in Eq.~(\ref{eq:ratio_th}) and use the experimentally 
measured values of the $B \to K^* \gamma$ branching fractions 
from Table~\ref{tab:exp-data}. A number of input hadronic quantities
has been changed compared to our earlier analysis~\cite{Ali:2004hn}
and the changes are desribed below.

Let us start with the discussion 
of the tensor $B \to V$ transition form factors. 
The $SU (3)_{\rm F}$-breaking effects in the QCD transition form 
factors $T_1^{(K^*)}(0)$, $T_1^{(\rho)}(0)$, and $T_1^{(\omega)}(0)$ 
have been evaluated in a number of different theoretical frameworks. 
We take the $SU(3)_{\rm F}$-breaking to hold also for the ratio 
of the soft form factors in the effective theory. Defining 
$\zeta \equiv \xi_\perp^{(\rho)} (0)/\xi_\perp^{(K^*)} (0)$, 
and restricting ourselves to the  QCD LCSRs,
we note that the earlier result in this approach~\cite{Ali:1993vd}, 
yielding $\zeta = 0.76 \pm 0.06$, has been updated recently 
yielding $\zeta = 0.86 \pm 0.07$~\cite{Ball:2006nr}, 
which we use here for the numerical analysis.
In our paper~\cite{Ali:2004hn}, we had assumed the equality of
the tensor form factors in the decays  $B \to \rho\gamma$ 
and $B \to \omega\gamma$, which holds in the $SU(3)_{\rm F}$ symmetry
limit. Recent estimates within the QCD LSCRs result in modest 
$SU (3)_{\rm F}$-breaking effect in the form factors, illustrated by 
the values~\cite{Ball:2004rg}: $T_1^{(\rho)} (0) = 0.267 \pm 0.021$ 
and $T_1^{(\omega)}(0) = 0.242 \pm 0.022$. This gives for  
 the ratio $\zeta_{\omega/\rho} \equiv 
\xi_\perp^{(\omega)}(0)/\xi_\perp^{(\rho)}(0)=0.9 \pm 0.1$, which in turn  
yields
$\xi_\perp^{(\omega)} (0)/\xi_\perp^{(K^*)} (0) = 
\zeta \, \zeta_{\rho/\omega} = 0.78 \pm 0.10$. This is  
 used in the analysis of the $B \to \omega\gamma$ decay.

We now discuss the changes connected with the  $B$-, 
$\rho$- and $K^*$-meson distribution amplitudes. In our earlier
paper~\cite{Ali:2004hn}, 
the two-parameter model for the leading-twist $B$-meson 
LCDA by Braun, Ivanov and Korchemsky (BIK)~\cite{Braun:2003wx} 
was used with the following ranges of the parameters: 
$\lambda_B^{-1} (1~{\rm GeV}) = (2.15 \pm 0.50)$~GeV$^{-1}$ 
and $\sigma_B (1~{\rm GeV}) = 1.4 \pm 0.4$, obtained from 
the sum-rules analysis. Recently, Lee and Neubert~\cite{Lee:2005gz}
 have derived 
model-independent properties of the $B$-meson LCDA, obtaining 
explicit expressions for the first two moments as a 
function of the renormalization scale~$\mu$. 
Based on this analysis, these authors suggest a modified
leading-twist $B$-meson LCDA which is consistent with the 
moment relations. It was also shown that the BIK model obeys 
the same moment constraints with the modified values of the
two input parameters
$\lambda_B^{-1} (1~{\rm GeV}) = (1.79 \pm 0.06)$~GeV$^{-1}$        
and $\sigma_B (1~{\rm GeV}) = 1.57 \pm 0.27$~\cite{Lee:2005gz}.
Though the functional forms of the two B-meson LCDAs are
different, with the indicated values of $\lambda_B^{-1}$
and $\sigma_B$,
both the BIK and the Lee-Neubert functions are
nearly indistinguishable. Following this work, we use 
the BIK model with the improved parameters in our analysis. 
For the $B$-meson decay constant, the value 
$f_B = (205 \pm 25)$~MeV~\cite{Ball:2006nr} is taken. 
The $\rho$-meson leading-twist LCDA was taken from 
Ref.~\cite{Ball:2006nr} with $f_\perp^{(\rho)} (1~{\rm GeV}) 
= (165 \pm 9)$~MeV and $a_{\perp 2}^{(\rho)} (1~{\rm GeV}) = 
0.15 \pm 0.07$. The models for the leading-twist LCDAs of the 
$K$- and $K^*$-meson have been updated during the last several 
years. In the present analysis we use the set of parameters 
for the $K^*$-meson LCDA from Ref.~\cite{Ball:2006nr}: 
$f_\perp^{(K^*)} (1~{\rm GeV}) = (185 \pm 10)$~MeV, 
$a_{\perp 1}^{(K^*)} (1~{\rm GeV}) = 0.04 \pm 0.03$, and 
$a_{\perp 2}^{(K^*)} (1~{\rm GeV}) = 0.11 \pm 0.09$. 
While~$f_\perp^{(K^*)}$ and~$a_{\perp 2}^{(K^*)}$ 
remain approximately the same, the first Gegenbauer 
moment~$a_{\perp 1}^{(K^*)}$ has changed significantly
from its previously used value, 
  $a_{\perp 1}^{(K^*)} (1~{\rm GeV}) = 
-0.34 \pm 0.18$. 
Note that the soft part, $\xi_\perp^{(K^*)}$, of the QCD 
form factor~$T_1^{K^*} (0)$, is practically insensitive 
to the changes in the $K^*$-meson LCDA, and the updated 
value now is $\bar \xi_\perp^{(K^*)} (0) = 0.26 \pm 0.02$.

The other sizable changes compared to our previous 
analysis~\cite{Ali:2004hn} are in the values of the CKM parameters,
which are now input. Taking into account the recent measurement of the 
ratio~$|V_{td}/V_{ts}|$
from the ratio~$\Delta M_d/\Delta M_s$ of the $B^0_d$- 
and $B^0_s$-meson mass differences by the CDF 
collaboration~\cite{:2006ze}, yields $|V_{td}/V_{ts}| = 
0.2060^{+0.0081}_{-0.0060} ({\rm theory}) \pm 0.0007 ({\rm exp})$, 
which we take as $| V_{td}/V_{ts}| = 0.206 \pm 0.008$. In addition,
the numerical value of the 
unitarity-triangle angle~$\alpha = (97.3^{+4.5}_{-5.0})^\circ$ is
taken from the global CKM fits~\cite{Charles:2004jd}. We also modify the
top quark mass, reflecting  
 the smaller value of the $t$-quark mass reported recently 
by the Fermilab collider experiments
$m_t = (171.4 \pm 2.1)$~GeV~\cite{Heinson:2006yq}.

The main uncertainties in the dynamical functions $\Delta R(\rho/K^*)$ 
and $\Delta R(\omega/K^*)$ come from the  CKM 
angle~$\alpha$ and the soft form factors~$\xi^{(K^*)}_\perp (0)$,  
$\xi^{(\rho)}_\perp (0)$, and~$\xi^{(\omega)}_\perp (0)$.  
Taking into account various parametric uncertainties, it is found that 
the dynamical functions are constrained in the ranges: 
% ranges~\cite{Ali:2001ez}:
%
\begin{equation}
\Delta R(\rho^\pm/K^{*\pm}) = 0.057^{+0.057}_{-0.055} ,
\quad
\Delta R(\rho^0/K^{*0}) = 0.006^{+0.046}_{-0.043} , 
\quad
\Delta R(\omega/K^{*0}) = -0.002^{+0.046}_{-0.043} .  
\end{equation}
Thus, these corrections turn out to be below~5\% in the  
radiative decays of the neutral $B$-meson, and may reach 
as high as~11\% for the charged mode. 
This explicitly quantifies the statement 
that the ratios $R_{\rm th} (\rho \gamma/K^* \gamma)$ and 
$R_{\rm th} (\omega \gamma/K^* \gamma)$~(\ref{eq:ratio_th}) 
are stable against~$O(\alpha_s)$ and $1/m_b$-corrections, 
in particular for the neutral $B$-meson decays. 
Comparison with the corresponding estimates obtained 
by us in Ref.~\cite{Ali:2004hn} shows that the central 
values are now smaller and the errors have decreased 
due to the various improvements since then. 
Note that the reduced central values reflect mainly   
the substantial change in the value of the input
parameter~$a_{\perp 1}^{(K^*)}$.   

With the modified input values specified above, the branching ratios for the
 radiative $B$-decays are estimated as follows: 
\begin{eqnarray}
&& \bar {\cal B}_{\rm th} (B^\pm \to \rho^\pm \gamma) =
(1.37 \pm 0.26 [{\rm th}] \pm 0.09 [{\rm exp}]) \times 10^{-6},
\nonumber \\
&& \bar {\cal B}_{\rm th} (B^0 \to \rho^0 \gamma) =
(0.65 \pm 0.12 [{\rm th}] \pm 0.03 [{\rm exp}]) \times 10^{-6},
\label{eq:BR-theory} \\ 
&& \bar {\cal B}_{\rm th} (B^0 \to \omega \gamma) =
(0.53 \pm 0.12 [{\rm th}] \pm 0.02 [{\rm exp}]) \times 10^{-6}~.
\nonumber
\end{eqnarray}
 In the above estimates, the first error is due 
to the uncertainties of the theory and the second error is 
from the experimental data on the $B \to K^* \gamma$ branching 
fractions. The recent data from the BABAR and BELLE experiments 
are in the right ball-park compared to the above SM-based 
predictions. However, the comparison of theory and experiment 
is not yet completely quantitative due to the paucity of data.

Combining all the above branching fractions~(\ref{eq:BR-theory}) 
together into the isospin- and $SU (3)_{\rm F}$-averaged branching 
fraction~(\ref{eq:BR-average-def}), one has the following prediction: 
\begin{equation} 
\bar {\cal B}_{\rm th} [B \to (\rho/\omega) \gamma] =
(1.32 \pm 0.26) \times 10^{-6},
\label{eq:BR-isospin-theory} 
\end{equation}
in agreement with the current world average
(see Table~\ref{tab:exp-data}). 

The results~(\ref{eq:BR-theory}) can be compared with the 
predictions obtained within the pQCD approach~\cite{Lu:2005yz}: 
\begin{eqnarray} 
\bar {\cal B}_{\rm th} (B^\pm \to \rho^\pm \gamma) & = & 
(2.5 \pm 1.5) \times 10^{-6},
\nonumber \\ 
\bar {\cal B}_{\rm th} (B^0 \to \rho^0 \gamma) & = &  
(1.2 \pm 0.7) \times 10^{-6},
\label{eq:BR-pQCD} \\  
\bar {\cal B}_{\rm th} (B^0 \to \omega \gamma) & = & 
(1.1 \pm 0.6) \times 10^{-6}.   
\nonumber  
\end{eqnarray} 
The central values and the errors in the pQCD approach are 
typically a factor of two larger than the improved QCDF-based 
predictions given earlier. An updated analysis of these branching
ratios in the pQCD approach will shed light on the current numerical
differences.

\paragraph{Direct CP-Asymmetry:}
The direct CP-asymmetry in the $B^\pm \to \rho^\pm \gamma$ 
decays is defined as follows: 
\begin{equation}
{\cal A}_{\rm CP} (\rho^\pm \gamma) =
\frac{{\cal B} (B^- \to \rho^- \gamma) - {\cal B} (B^+ \to \rho^+ \gamma)}
  {{\cal B} (B^- \to \rho^- \gamma) + {\cal B} (B^+ \to \rho^+ \gamma)} . 
\label{eq:CP-asymmetry}
\end{equation}
In NLO, the direct CP-asymmetry 
can be written in the form~\cite{Ali:2001ez,Ali:2000zu}: 
\begin{equation} 
{\cal A}_{\rm CP} (\rho^\pm \gamma) = 
\frac{2 \vert \lambda_u\vert \sin \alpha}
     {C_7^{(0) {\rm eff}} \, (1 + \Delta_{\rm LO})} \, 
{\rm Im} \left [ A^u - \epsilon_A \ A^{(1)t} \right ] , 
\label{eq:ACP-dir-NLO}
\end{equation}
where $\lambda_u$  has been defined in Eq.~(\ref{eq:CKM-ratio})
and $\Delta_{\rm LO}$ is the isospin-violating ratio in 
the leading order~\cite{Ali:2001ez,Ali:2000zu}: 
\begin{equation} 
\Delta_{\rm LO} = - 2 \epsilon_A \, |\lambda_u| \cos \alpha 
+ \epsilon_A^2 \, |\lambda_u|^2 .     
\label{eq:Delta-LO} 
\end{equation}
 
\begin{figure}[tb]
\centerline{\epsfxsize=.37\textwidth
            \epsffile{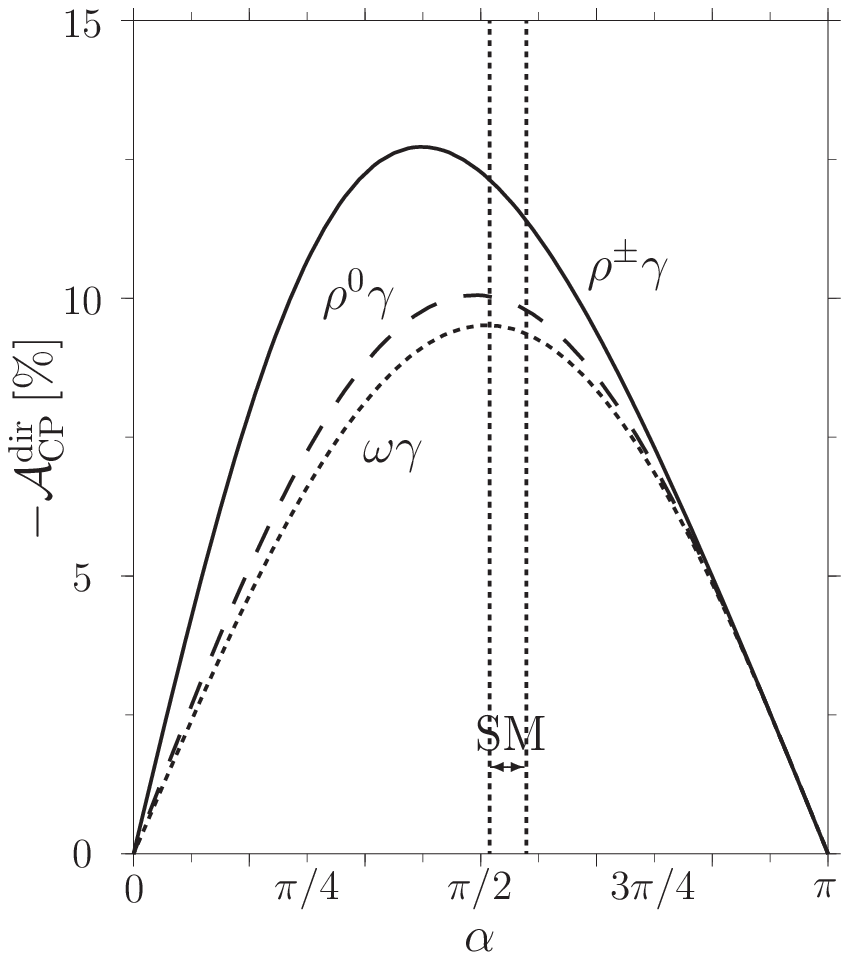} \qquad
            \epsfxsize=.37\textwidth
            \epsffile{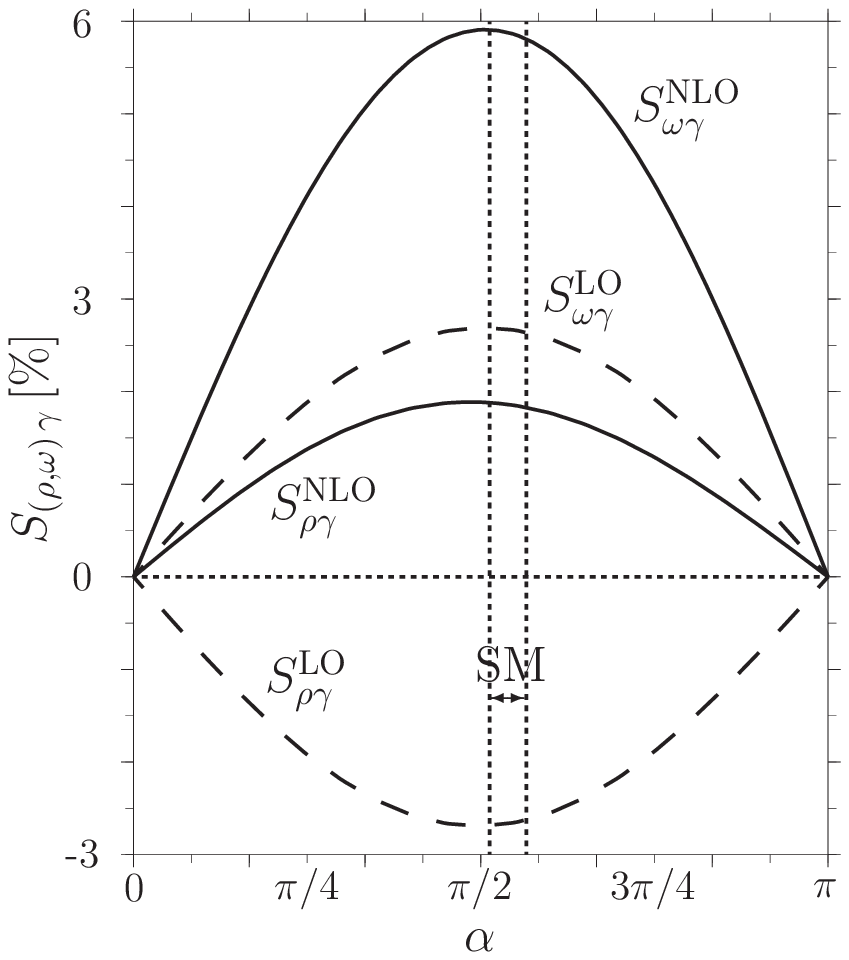}}
\caption{Left figure: Direct CP-asymmetry in the decays 
         $B^\pm \to \rho^\pm \gamma$ (solid curve), 
         $B^0 \to \rho^0 \gamma$ (dashed curve) and 
         $B^0 \to \omega \gamma$ (dotted curve) as a function 
         of the unitarity-triangle angle~$\alpha$. 
         Right figure: Mixing-induced CP-asymmetry in the decays
         $B^0 \to \rho^0 \gamma$ (solid curves) and 
         $B^0 \to \omega \gamma$ (dotted curves) in the leading 
         (LO) and next-to-leading (NLO) orders as a function 
         of the unitarity-triangle angle~$\alpha$.  
         The $\pm 1 \sigma$ allowed band of~$\alpha$ from 
         the SM unitarity fits~\cite{Charles:2004jd} 
         is also indicated on both plots.}
\label{fig:ACPdir}
\end{figure}

Similar definitions and expressions can also be used for the two neutral 
decay modes
$B^0 \to \rho^0\gamma$ and $B^0 \to \omega\gamma$. 
The dependence of the CP-asymmetry on the angle~$\alpha$ 
for the three decay modes 
is presented in the left plot in Fig.~\ref{fig:ACPdir}. 
%
%%%%%%%%%%%%%%%%%%%%%%%%%%%%%%%%%%%%%%%%%%%%%%%%%%%%%%%%%%%%%%%%%%%%%%%%%
%
%\hspace{90mm}
%
%%%%%%%%%%%%%%%%%%%%%%%%%%%%%%%%%%%%%%%%%%%%%%%%%%%%%%%%%%%%%%%%%%%%%%%%%
%
In the QCDF approach, the SM  yields the direct CP-asymmetry 
to be negative, and the results in the interval 
$0.21 \le \sqrt z = m_c/m_b \le 0.33$ are as follows:
\begin{equation} 
{\cal A}_{\rm CP} (\rho^\pm \gamma) = 
\left ( -11.8^{+2.8}_{-2.9} \right )\% , 
\quad 
{\cal A}_{\rm CP} (\rho^0 \gamma) = 
\left ( -9.9^{+3.8}_{-3.4} \right )\% , 
\quad 
{\cal A}_{\rm CP} (\omega \gamma) = 
\left ( -9.5^{+4.0}_{-3.6} \right )\% .  
\label{eq:ACPdir-SM-results}
\end{equation}
Being at the level of~10\%, the direct CP asymmetry
in these decays can be measured at the current 
$B$-factories in several years. 

The results~(\ref{eq:ACPdir-SM-results}) can be compared with the 
predictions obtained within the pQCD approach~\cite{Lu:2005yz}: 
\begin{equation} 
{\cal A}_{\rm CP} (\rho^\pm \gamma) = 
\left ( 17.7 \pm 15.0 \right )\% , 
\quad 
{\cal A}_{\rm CP} (\rho^0 \gamma) = 
\left ( 17.6 \pm 15.0 \right )\% , 
\quad 
{\cal A}_{\rm CP} (\omega \gamma) = 
\left ( 17.9 \pm 15.2 \right )\% .  
\label{eq:ACPdir-pQCD-results}
\end{equation}
They are at variance with the results~(\ref{eq:ACPdir-SM-results}) 
based on the QCD factorization discussed here. In particular, the 
direct CP-asymmetry is predicted to be positive in the pQCD approach 
in all decay modes and the central values are typically a factor 
of two larger while errors are rather large.  Measurements of these 
asymmetries will allow to distinguish the detailed dynamical models 
illustrated here by the differing predictions of the QCDF and pQCD 
approaches.

\paragraph{Mixing-Induced CP-Asymmetry:}
For the time-dependent CP-asymmetries in the neutral $B$-meson 
decay modes, the interference of the $B^0 - \bar B^0$-mixing 
and decay amplitudes has to be taken into account, yielding the 
following characteristic time-dependence of such asymmetries: 
\begin{eqnarray} 
a_{\rm CP}^{\rho \gamma} (t) & = &  
- C_{\rho \gamma} \cos (\Delta M_d \, t) 
+ S_{\rho \gamma} \sin (\Delta M_d \, t) ,   
\label{eq:ACP-time-rho} \\ 
a_{\rm CP}^{\omega \gamma} (t) & = &  
- C_{\omega \gamma} \cos (\Delta M_d \, t) 
+ S_{\omega \gamma} \sin (\Delta M_d \, t) ,   
\label{eq:ACP-time-omega}  
\end{eqnarray} 
where $\Delta M_d$ is the $B_d^0 - \bar B_d^0$
mass difference. The coefficients~$C_{\rho \gamma}$ 
and~$C_{\omega \gamma}$ accompanying $\cos (\Delta M_d \, t)$ 
in Eqs.~(\ref{eq:ACP-time-rho}) and~(\ref{eq:ACP-time-omega}),  
up to a sign, coincide with the direct CP-asymmetry discussed above. 
The second coefficients~$S_{\rho \gamma}$ and $S_{\omega \gamma}$,
called the mixing-induced CP-asymmetries, are defined as follows: 
\begin{eqnarray} 
S_{\rho \gamma} = \frac{2 \, {\rm Im} (\lambda_{\rho \gamma})}
                       {1 + |\lambda_{\rho \gamma}|^2},  
\quad  
\lambda_{\rho \gamma} \equiv \frac{q}{p} \, 
\frac{A (\bar B^0 \to \rho^0 \gamma)}{A (B^0 \to \rho^0 \gamma)} , 
\label{eq:ACP-mix-rho-def} \\ 
S_{\omega \gamma} = \frac{2 \, {\rm Im} (\lambda_{\omega \gamma})}
                       {1 + |\lambda_{\omega \gamma}|^2},  
\quad  
\lambda_{\omega \gamma} \equiv \frac{q}{p} \, 
\frac{A (\bar B^0 \to \omega \gamma)}{A (B^0 \to \omega \gamma)} , 
\label{eq:ACP-mix-omega-def} 
\end{eqnarray} 
where the ratio~$q/p = {\rm e}^{- 2 i \beta}$ 
is a pure phase factor to a good accuracy
(experimentally, $|q/p| = 1.0013 \pm 0.0034$~\cite{HFAG}). 

The mixing-induced CP-violating asymmetry~$S_{\rho\gamma}$ 
in NLO can be presented in the form~\cite{Ali:2004hn}:  
\begin{equation}
S_{\rho \gamma}^{\rm LO} = -   
\frac{2 |\lambda_u| \, \varepsilon_A^{(0)} \sin \alpha \, 
     (1 - |\lambda_u| \, \varepsilon_A^{(0)} \cos \alpha)}
     {1 - 2 |\lambda_u| \, \varepsilon_A^{(0)} \cos \alpha + 
      |\lambda_u|^2 (\varepsilon_A^{(0)})^2} ,   
\label{eq:S(rho-gam)-LO}    
\end{equation} 
\begin{equation} 
S_{\rho \gamma}^{\rm NLO} =  
S_{\rho \gamma}^{\rm LO} - \frac{2 |\lambda_u| \sin\alpha \,  
[1 - 2 |\lambda_u| \, \varepsilon_A^{(0)} \cos\alpha +  
|\lambda_u|^2 (\varepsilon_A^{(0)})^2 \cos (2\alpha)]} 
{[1 - 2 |\lambda_u| \, \varepsilon_A^{(0)} \cos\alpha + 
|\lambda_u|^2 (\varepsilon_A^{(0)})^2]^2} \, 
\frac{A_R^u - \varepsilon_A^{(0)} A_R^{(1) t}}{C_7^{(0) {\rm eff}}} , 
\label{eq:S(rho-gam)-NLO}  
\end{equation} 
where $A_R^{(1)t}$ and $A_R^u$ are the real parts 
of the NLO contributions to the decay amplitudes.  
This expression can be easily rewritten for~$S_{\omega \gamma}$. 
It is seen that, neglecting the weak-annihilation 
contribution ($\varepsilon_A^{(0)} = 0$), 
the mixing-induced CP-asymmetry vanishes in the leading 
order. However, including the $O(\alpha_s)$ contribution, 
this CP-asymmetry is non-zero even
in the absence of the annihilation contribution.   
The dependence of the mixing-induced CP-asymmetry 
on the angle~$\alpha$ is presented in the right plot 
in Fig.~\ref{fig:ACPdir}.  

The QCDF-based estimates of the mixing-induced CP-asymmetry 
in the leading and next-to-leading order in~$\alpha_s$ in the SM are: 
\begin{eqnarray} 
S_{\rho \gamma}^{\rm LO}  = (-2.7 \pm 0.9)\% , 
\qquad 
S_{\rho \gamma}^{\rm NLO} = (1.9^{+3.8}_{-3.2})\% ,  
\label{eq:ACP-mix-results} \\
S_{\omega \gamma}^{\rm LO}  = (+2.7 \pm 0.9)\% , 
\qquad 
S_{\omega \gamma}^{\rm NLO} = (5.9^{+4.1}_{-3.5})\% , 
\nonumber 
\end{eqnarray} 
showing the tendency of the NLO corrections to compensate 
the leading order contribution in~$S_{\rho \gamma}$ and 
enhancing it in~$S_{\omega \gamma}$.  
Theoretical uncertainties are rather large and both the values 
are consistent with being small.

\begin{figure}[bt]
\centerline{\epsfysize=.37\textwidth 
            \epsffile{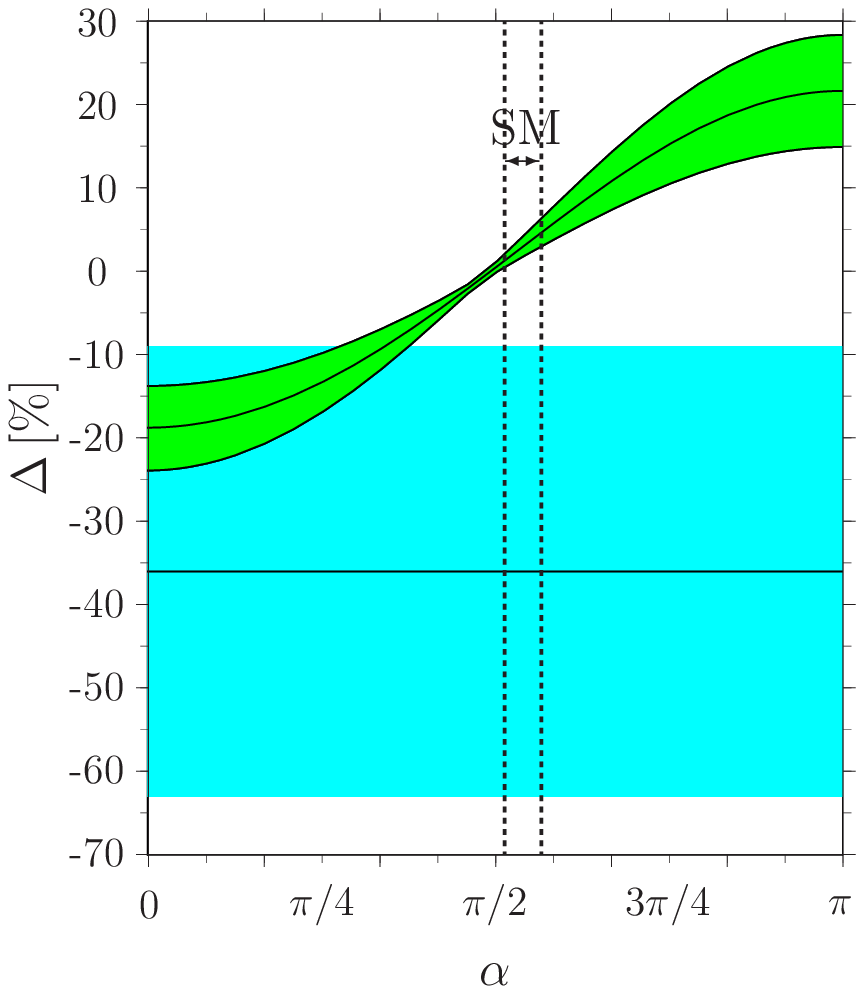} \qquad 
            \epsfysize=.37\textwidth  
            \epsffile{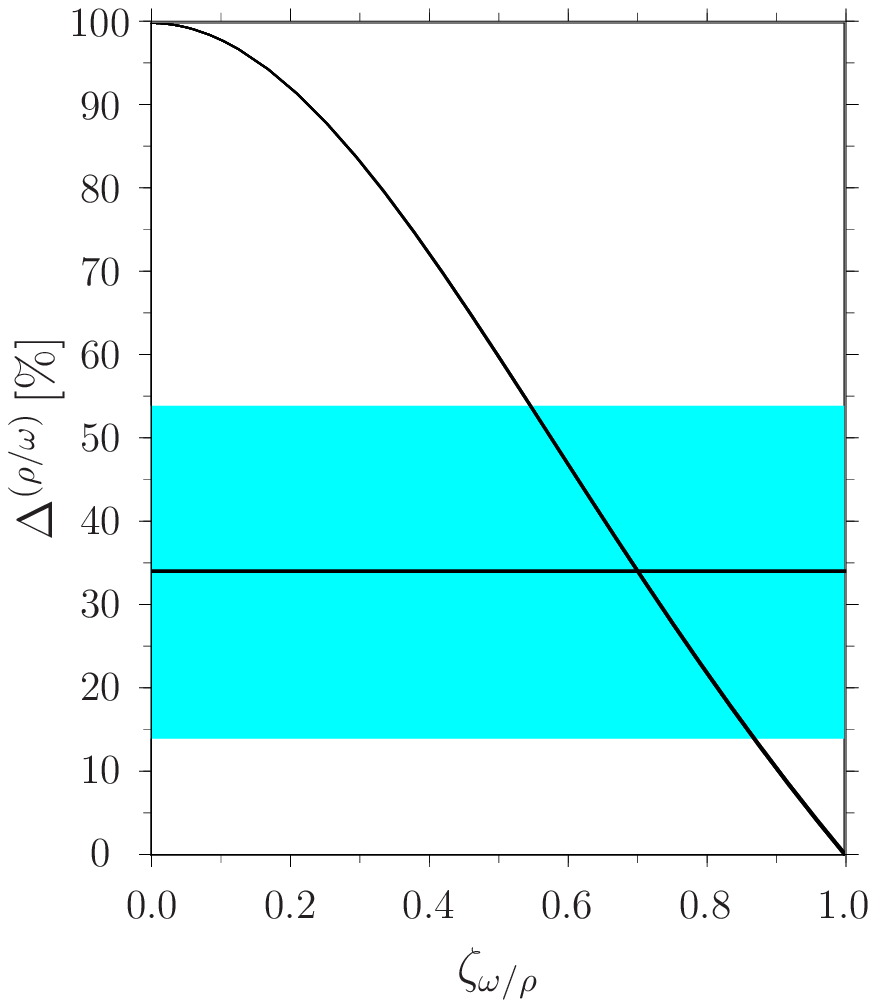}}
\caption{Left figure: 
         The charged-conjugate averaged ratio~$\Delta$ 
         for $B \to \rho \gamma$ decays (green/dark-shaded region)   
         as a function of the unitarity-triangle angle~$\alpha$. 
         The blue/shaded area is the experimentally measured  
         region by the BABAR collaboration~\cite{unknown:2006ag}:
         $\Delta_{\rm exp} = -0.36 \pm 0.27$.
         Right figure: 
         The $SU (3)_{\rm F}$-violating ratio~$\Delta^{(\rho/\omega)}$ 
         as a function of the ratio~$\zeta_{\omega/\rho} = 
         \xi_\perp^{(\omega)}(0)/\xi_\perp^{(\rho)}(0)$.  
         The blue/shaded area is the experimental region: 
         $\Delta^{(\rho/\omega)}_{\rm exp} = 0.34 \pm 0.20$, 
         determined from the HFAG averages~\cite{HFAG}. 
} 
\label{fig:Delta}
\end{figure}

\paragraph{Isospin-Violating Ratio:} 
The charge-conjugate averaged quantity~$\Delta$ 
for the $B \to \rho \gamma$ decays defined as:
\begin{equation}
\Delta = \frac{1}{4} \,
\left [
\frac{\Gamma (B^- \to \rho^- \gamma)}
     {\Gamma (\bar B^0 \to \rho^0 \gamma)}
+ \frac{\Gamma (B^+ \to \rho^+ \gamma)}
       {\Gamma (B^0 \to \rho^0 \gamma)}
\right ] - 1 ,
\label{eq:Delta}
\end{equation}
is found to be stable against the NLO and
$1/m_b$-corrections~\cite{Ali:2001ez}. 
In the leading-order, this ratio has
been defined in Eq.~(\ref{eq:Delta-LO}).
The NLO corrections do not change the LO result significantly 
and preserve the main feature~-- the small value in the  
vicinity of $\alpha = 90^\circ$, the region favored 
by the CKM fits~\cite{Charles:2004jd,Bona:2006ah}.  
The dependence of the isospin-violating ratio 
on the angle~$\alpha$ is presented in the left plot in 
Fig.~\ref{fig:Delta}.  
%
%%%%%%%%%%%%%%%%%%%%%%%%%%%%%%%%%%%%%%%%%%%%%%%%%%%%%%%%%%%%%%%%%%%%%%%%%
%
%\hspace{90mm}
%
%%%%%%%%%%%%%%%%%%%%%%%%%%%%%%%%%%%%%%%%%%%%%%%%%%%%%%%%%%%%%%%%%%%%%%%%%
%
In the expected ranges of the CKM parameters~\cite{Charles:2004jd}, 
this ratio is estimated as $\Delta = (2.9 \pm 2.1)\%$. A~comparison 
with the result obtained within the pQCD approach~\cite{Lu:2005yz}: 
$\Delta = (-5.4 \pm 5.4)\%$, shows that, apart from being somewhat 
larger in magnitude, $\Delta$ has the opposite sign. 
Thus, while~$\epsilon_A$ is model-dependent, explicit calculations 
show that the SM predicts a small isospin-violation in the 
$B \to \rho\gamma$ decays, as its measure,~$\Delta$, is parametrically 
suppressed (being proportional to $\cos\alpha$, with~$\alpha$ close 
to~$90^\circ$). A comparison of~$\Delta$ with the recent BABAR 
measurement~(\ref{eq:Delta-exp}) of the same is shown in
Fig.~\ref{fig:Delta}. As the current experimental errors are
rather large, one  will have to wait for higher statistics data from the 
$B$-factories to draw any quantitative conclusion.

\paragraph{\boldmath{$SU (3)_{\rm F}$}-Violating Ratio:} 
The ratio based on the branching fractions of the neutral 
$B^0 \to \rho^0\gamma$ and $B^0 \to \omega\gamma$ decay modes 
may be defined as follows: 
\begin{eqnarray}  
\Delta^{(\rho/\omega)} & \equiv & \frac{1}{2} \left [ 
\Delta^{(\rho/\omega)}_B + \Delta^{(\rho/\omega)}_{\bar B} \right ] ,  
\label{eq:SU3-ratio} \\ 
\Delta^{(\rho/\omega)}_B & \equiv &   
\frac{(m_B^2 - m_\omega^2)^3 \, {\cal B} (B^0 \to \rho^0 \gamma) - 
      (m_B^2 - m_\rho^2)^3 \, {\cal B} (B^0 \to \omega \gamma)}
     {(m_B^2 - m_\omega^2)^3 \, {\cal B} (B^0 \to \rho^0 \gamma) + 
      (m_B^2 - m_\rho^2)^3 \, {\cal B} (B^0 \to \omega \gamma)} .   
\nonumber 
\end{eqnarray} 
The NLO expression obtained in the $SU (3)_{\rm F}$ symmetry limit, 
$\zeta_\perp^{(\rho)} (0) = \zeta_\perp^{(\omega)} (0)$, 
can be written in a simple form~\cite{Ali:2004hn}: 
\begin{equation} 
\Delta^{(\rho/\omega)}_{SU(3)} = -  
\frac{|\lambda_u| \, (\varepsilon_A^{(0)} - \varepsilon_A^{(\omega)})}
     {C_7^{(0) {\rm eff}}} 
\left [ (C_7^{(0) {\rm eff}} - A_R^{(1) t}) \cos\alpha 
+ |\lambda_u| \,  A_R^u \cos (2\alpha) \right ] .    
\label{eq:Delta-rho/omega-approx}
\end{equation} 
The theoretical expression~(\ref{eq:Delta-rho/omega-approx}) 
for the ratio~$\Delta^{(\rho/\omega)}$ can be improved 
by including the $SU(3)_{\rm F}$-breaking in the ratio~$\zeta_{\omega/\rho}$:  
\begin{equation} 
\Delta^{(\rho/\omega)} = 
\frac{1 - \zeta_{\omega/\rho}^2}{1 + \zeta_{\omega/\rho}^2} 
+ \frac{4 \zeta_{\omega/\rho}^2}{(1 + \zeta_{\omega/\rho}^2)^2} \,  
\Delta^{(\rho/\omega)}_{SU(3)} + 
{\cal O} (\alpha_s^2, \varepsilon_A^{(0)} \varepsilon_A^{(\omega)}).       
\label{eq:Delta-rho/omega-complete} 
\end{equation} 
The dependence of~$\Delta^{(\rho/\omega)}$ 
on the parameter~$\zeta_{\omega/\rho}$ is presented 
in the right plot in Fig.~\ref{fig:Delta}. 
Based on the recent averages from Table~\ref{tab:exp-data}, 
one obtains the following experimental estimate: 
$\Delta^{(\rho/\omega)}_{\rm exp} = (34 \pm 20)\%$, which is
also shown in Fig.~\ref{fig:Delta}. 
Within the range $\zeta_{\omega/\rho} = 0.9 \pm 0.1$, 
derived from the results of Ref.~\cite{Ball:2004rg}, 
we estimate: 
$\Delta^{(\rho/\omega)} = (11 \pm 11)\%$, which 
is consistent with the experimental value within  large errors. 
We remark that $\Delta^{(\rho/\omega)}$  is dominated by the first term
in Eq.~(\ref{eq:Delta-rho/omega-complete}), as its $SU(3)_{\rm F}$-
symmetric value $\Delta^{(\rho/\omega)}_{SU(3)}$ is estimated to be small,
$\Delta^{(\rho/\omega)}_{SU(3)}=(2.0 \pm 1.9) \times 10^{-3}$.

%
%%%%%%%%%%%%%%%%%%%%%%%%%%%%%%%%%%%%%%%%%%%%%%%%%%%%%%%%%%%%%%%%%%%%%%%%%
%
%\hspace{90mm}
\begin{figure}[bt]
\centerline{\epsfysize=.40\textwidth 
            \epsffile{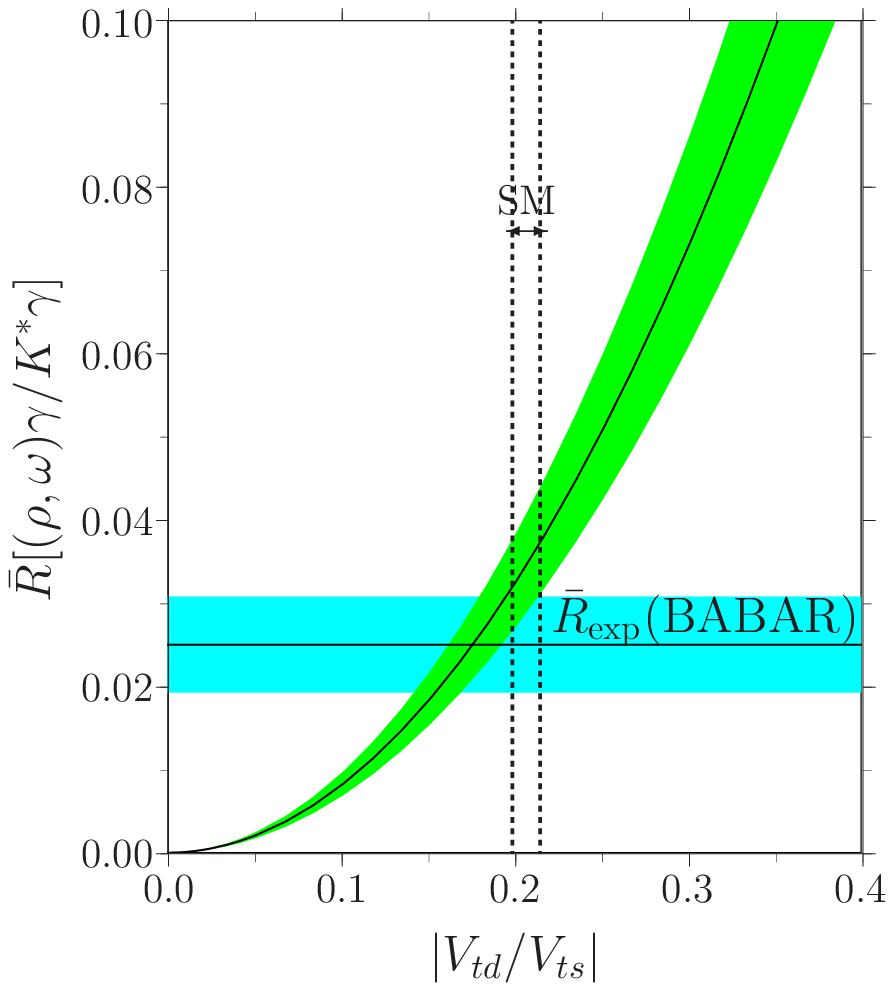} \qquad 
            \epsfysize=.40\textwidth 
            \epsffile{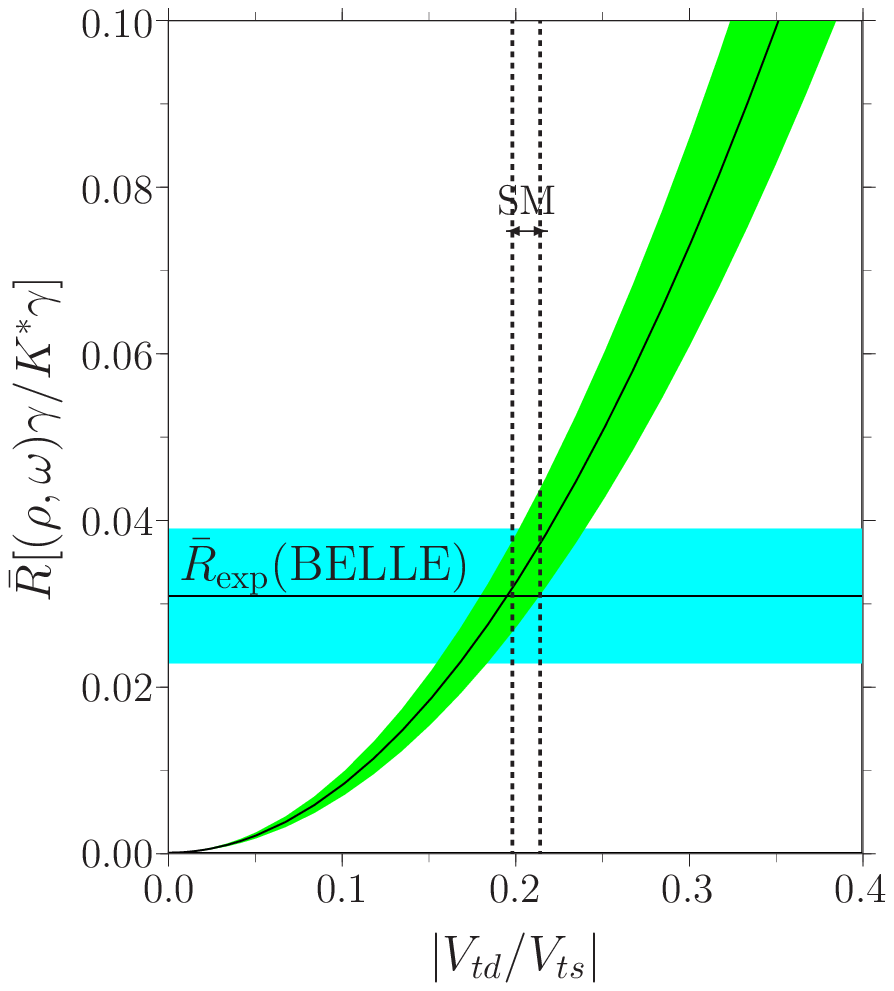}}
\caption{The ratio~$\bar R[(\rho,\omega)\gamma/K^*\gamma]$ 
         as a function of the ratio~$|V_{td}/V_{ts}|$ of the 
         CKM matrix elements. The plots based on the BABAR 
         (left figure) and BELLE (right figure) measurements 
         of the isospin-averaged $B \to (\rho,\omega) \gamma$ 
         branching fractions show their good agreement both 
         with the theoretical estimations of this ratio (green 
         region) and with the recent CDF measurement~\cite{:2006ze} 
         (the vertical band labeled as~``SM'') within the~$1\sigma$ 
         intervals.}   
\label{fig:Ratio}
\end{figure}
%
%%%%%%%%%%%%%%%%%%%%%%%%%%%%%%%%%%%%%%%%%%%%%%%%%%%%%%%%%%%%%%%%%%%%%%%%%
%

\section{Summary} 

Physics of the radiative $B \to \rho\gamma$ and $B \to \omega\gamma$ 
decays will impact on the CKM phenomenology. A good measure of this
is the value of the CKM ratio $|V_{td}/V_{ts}|$, which can be extracted
from these decays in conjunction with the $B \to K^*\gamma$ decays.
First results along these lines have been obtained by the BABAR and 
BELLE collaborations, which will become quantitative in due course 
of time. In addition to the branching fractions resulting in the 
estimates of the ratio~$|V_{td}/V_{ts}|$ (see Fig.~\ref{fig:Ratio}), 
the analysis of different asymmetries in these modes will give 
additional information on the CKM parameters, in particular on the 
unitarity-triangle angle~$\alpha$, apart from shedding light on the 
underlying QCD dynamics.  
In this review, we have taken the attitude that the CKM parameters are 
well known by now and we use this input to make definite predictions 
for the branching ratios and various related asymmetries in the 
$B \to (\rho,\omega)\gamma$ decays. The SM-based predictions are 
in fair agreement with data and this comparison will become more 
precise in the coming years.

\section*{Acknowledgments}

A.P. would like to express his deep gratitude 
to the organizers of the International Seminar ``Quarks-2006'' 
for their warm hospitality, 
and thank the Theory Group at DESY for their kind hospitality, 
where a part of this paper was written. 
The financial support in the framework of the ``Michail Lomonosov'' 
Program by the German Academic Exchange Service (DAAD) and 
the Ministry of Education and Science of the Russian Federation
is also gratefully acknowledged. 
This work was supported in part by the Council on Grants 
by the President of the Russian Federation for the Support 
of Young Russian Scientists and Leading Scientific Schools 
of the Russian Federation under the Grant No. NSh-6376.2006.2, 
and by the Russian Foundation for Basic Research 
under the Grant No. 04-02-16253.

\end{document}